\documentclass[12pt]{article}
\usepackage[latin1]{inputenc}
\usepackage{slashed}
\usepackage{amsmath}
\usepackage{textcomp}
\usepackage{amssymb}
\usepackage{amsfonts}
\usepackage{indentfirst}
\usepackage{color}
\usepackage[top=2cm,bottom=2cm,left=3cm,right=2cm]{geometry}
\newcommand{\nin}{\noindent}
\newcommand{\be}{\begin{equation}}
\newcommand{\ee}{\end{equation}}
\newcommand{\bea}{\begin{eqnarray}}
\newcommand{\eea}{\end{eqnarray}}

\newcommand{\nn}{\nonumber\\}

\newcommand{\ol}{\overline}

\usepackage{graphicx}

\begin{document}

KCL-PH-TH/2015-13

\begin{center} 
 
 {\Large{\bf Effective fermion kinematics \\from modified quantum gravity}}

\vspace{0.5cm}

{\bf J. Alexandre} and {\bf J. Leite}

\vspace{0.2cm}

King's College London, Department of Physics, WC2R 2LS, UK\\
{\small jean.alexandre@kcl.ac.uk, julio.leite@kcl.ac.uk}

\vspace{1cm}

{\bf Abstract}

\end{center}

\vspace{0.5cm}

We consider a classical fermion and a classical scalar, propagating on two different kinds of 4-dimensional diffeomorphism breaking gravity backgrounds, 
and we derive the one-loop effective dispersion relation for matter, after integrating out gravitons. 
One gravity model involves quadratic divergences at one-loop, as in Einstein gravity, and the other model is the $z=3$ non-projectable Horava-Lifshitz
gravity, which involves logarithmic divergences only. 
Although these two models behave differently in the UV, the IR phenomenology for matter fields is comparable: {\it(i)} for generic values for the parameters, both models 
identify $10^{10}$ GeV as the characteristic scale above which they are not consistent with current upper bounds on Lorentz symmetry violation; {\it(ii)}
for both models, there is always a fine-tuning of parameters which allows the cancellation of the indicator for Lorentz symmetry violation.

\vspace{2cm}

\section{Introduction}

The high-energy behaviour of gravity is still not well understood, and several directions have been developed in order to build 
a satisfactory theoretical and phenomenological framework. One possibility is to consider gravity as an effective model \cite{donoghue} and study the 
implications of its quantisation at energies well below the Planck mass. Another possibility is to modify the ultraviolet (UV) behaviour of gravity, as can  
consistently be done, for example, with Horava-Lifshitz gravity \cite{Horava}.
We consider here two modifications of Einstein gravity which are invariant under the reduced symmetry of 3-dimensional diffeomorphisms, in the situation where a 
specific coordinate frame is preferred, and study the consequences of this violation of local Lorentz symmetry on the low energy physics.
In order to test the validity of these modified gravity models, we couple them to classical matter fields 
and study the generation, through quantum gravity corrections, of Lorentz-violating features in the matter dispersion relations.
These two models are
\begin{itemize}
\item A modification of Einstein gravity, which preserves only 3-dimensional space diffeomorphisms, as in  Horava-Lifshitz (HL) gravity, 
but which does not involve higher-order space derivatives or anisotropic scaling between space and time, so that one-loop corrections are quadratically divergent. This 
model is consistent with the effective approach to gravity; 
\item An extension of the first model, with an improved UV behaviour. This is achieved by the introduction  of higher-order 
space derivatives, leading to less divergent loop integrals, and which can consistently be implemented if we allow for an anisotropic scaling between space and time. The 
extension we consider is the non-projectable version of HL gravity (npHL) \cite{non-proj} for $z=3$, and involves logarithmic divergences only. 
\end{itemize}
In both cases, gravity is
naturally described in terms of the Arnowitt-Deser-Misner (ADM) decomposition of the metric, which naturally exhibits a space-time foliation.
Also, the graviton has 3 degrees of freedom (dof), with a scalar dof appearing as a consequence of the breaking of 4-dimensional diffeomorphisms \cite{Arkani}. 

We couple these models to classical matter and derive the one-loop effective dispersion relation seen by particles, after integrating out gravitons. One-loop quantum corrections to 
the dispersion relations of quantum matter fields coupled to HL gravity have been studied in~\cite{Pospelov}, where  the Authors derive 
the effective speed seen by a scalar field and an Abelian gauge field, and compare these to measure Lorentz-symmetry violation. In \cite{Padilla}, the non-projectable 
version of HL gravity is considered for the derivation of the effective matter Lagrangian. 
We are doing a similar study here, taking into account a classical scalar and fermionic backgrounds though, and we calculate the difference $\Delta v^2$ between the effective speeds 
of light seen by these two species.

The first model involves quadratic divergences, so that we use a cut off to calculate one-loop graviton loops,
whereas the second model involves logarithmic divergences only, that we control with dimensional regularisation. 
We find that, although the two models behave differently in the UV,
their phenomenological predictions are very similar in the sense that they predict the same order of magnitude for $\Delta v^2$, if one keeps generic values for the different parameters. 
Therefore this phenomenology  is not really improved by the introduction of higher-order space derivatives.
Moreover, in both cases, we find that one can always impose $\Delta v^2=0$ if the parameters are fine-tuned.

Studies of effective dispersion relations for Lifshitz-type models in flat space-time have shown the limitations of 
phenomenological viability of these models. This was first noted in \cite{IengoRussoSerone}, where the unnatural fine-tuning of bare parameters is shown 
in order to  match the light cones seen by two different scalar particles interacting. Similar studies were done in \cite{ABH}, 
where the effective dispersion relation for interacting Lifshitz fermions is derived, in the case where flavour symmetry is broken,
or in \cite{AB}, where the fermion effective dispersion relation is derived in a Lifshitz extension of Quantum Electrodynamics.
Our aim here is to study similar features and, more precisely, how global Lorentz-symmetry is affected from a local symmetry breaking.
A more complete study could involve the energy dependence of effective parameters obtained from a quantum modified gravity, in a Wilsonian renormalisation framework for example, as 
shown in \cite{Saueressig} for HL gravity. The latter approach is also used in \cite{Dodorico}, where instead the scalar field is integrated out on classical metric background.

Section 2 describes the models and discusses gauge freedom, as well as the actions for gravity and matter sectors. We describe the different simplifications which can be used 
for the one-loop integration of graviton, which is presented in section 3 and in the appendix. We integrate there successively the different components of graviton fluctuations, for both models, where
we keep on-shell the auxiliary fields which appear in the decomposition of the graviton fluctuations. 
Finally, we discuss the phenomenology of the two models in section 4, based on the current upper bounds for Lorentz-symmetry violating parameters.

\section{Models}

In this article, Greek letters denote space-time indices and Latin letters space indices only. The signature of the metric is $(-,+,+,+)$.

\subsection{The original actions}

We introduce here two modified gravity models, and their coupling to matter.\\

$\bullet$ The first modified gravity action we consider is a modified Einstein-Hilbert action where new operators are allowed
due to the reduced symmetry of the theory. Using the ADM decomposition and omitting the cosmological constant, it can be expressed as
\be\label{SG}
S_G = M_P^2 \int dt d^3 x \sqrt{g} N (K_{ij} K^{ij} -\lambda K^2 + R^{(3)}+\alpha a_i a^i)~,
\ee 
where $M_P^2 = (16 \pi G_N)^{-1}$ and
\bea 
g &=& |\det(g_{ij})| \\
K_{ij} &=& \frac{1}{2N}( \partial_t g_{ij} - D_i N_j - D_j N_i)~~~\mbox{with}~~~D_i N_j = \partial_i N_j - \Gamma^{k}_{ij} N_k \nn
K &=& K_{ij} g^{ij}~,~~~ R^{(3)} = R_{ijkl} g^{ik} g^{jl}~,~~~~a_i=\partial_i \ln N\nonumber~.
\eea 
$ G_N$ and $K_{ij}$ are respectively the Newton gravitational constant and the extrinsic curvature. 
The possibility to have $\lambda\ne1$ and $\alpha\neq 0$ implies that 4-dimensional diffeomorphisms are broken to 3-dimensional diffeomorphisms, and
the Hilbert-Einstein action can be recovered with $\lambda=1$ and $\alpha=0$.\\

$\bullet$ The second model we consider is the non-projectable version of Horava-Lifshitz gravity~\cite{non-proj} which allows the lapse 
function $N$ to be a function of space and time. This implies that terms containing $a_i = \partial_i\ln N$ should be included in the action, in order to have a 
consistent dispersion relation for the scalar graviton \cite{non-proj}, and avoids the well-known pathology present in the original HL gravity version~\cite{Horava}. 
In this context, we impose space and time to scale anisotropically such that
\be 
\vec{x}\to b\vec{x}~~~\mbox{while}~~~t\to b^z t~, 
\ee 
and the choice $z=3$ motivates us to introduce dimension 4 and dimension 6 operators, which become important in the ultraviolet (UV). As in~\cite{non-proj} though,
the time component is reparametrised so that the action is written in terms of the ``physical'' units.
The npHL action can be written as~\cite{non-proj}
\bea\label{SHL}
S_{HL} &=& M_P^2 \int dt d^3 x \sqrt{g} N \left\{K_{ij} K^{ij} -\lambda K^2 + R^{(3)}+\alpha a_i a^i\right.\\ 
&&\left. + F_1 R_{ij}R^{ij} +F_2 (R^{(3)})^2 +F_3 R^{(3)} \nabla_i a^i +F_4 a_i \Delta a^i\right.\nonumber\\
&&\left.+S_1 (\nabla_i R_{jk})^2 +S_2 (\nabla_i R^{(3)})^2 + S_3 (\Delta R^{(3)} \nabla_i a^i) + S_4 (a_i \Delta^2 a^i) \right\}~,\nonumber
\eea 
where $F_i= (f_i/M_{HL}^2)$ and $S_i=(s_i/M_{HL}^4)$ with $M_{HL}$ being the Horava-Lifshitz scale, and $f_i$ and $s_i$ are dimensionless coupling constants associated 
with operators of dimension 4 and 6, respectively.\\

$\bullet$ Finally, for the matter sector, we will consider complex scalar and fermion fields minimally coupled to the gravity models~(\ref{SG}) and (\ref{SHL}). The complex scalar field action is
\be\label{Ss1}
S_s = -\int dt d^3 x~\sqrt{g}Ng^{\mu\nu}\partial_\mu \phi \partial_\nu \phi^\star~,
\ee
and the fermion action is
\be\label{Sf1}
S_f = -\int dt d^3 x~\frac{ie}{2} \left[\bar{\psi} \gamma^\alpha e^{\mu}_{~\alpha} \nabla_\mu \psi-e^{\mu}_{~\alpha} 
(\nabla_\mu\bar{\psi}) \gamma^\alpha\psi \right]~,
\ee
where 
	\bea 
	e &=& \det(e^{~\alpha}_{\mu}) = \sqrt{g} N\\
	\nabla_\mu \psi &=& (\partial_\mu + \Gamma_\mu) \psi~~\mbox{and}~~ \nabla_\mu \bar{\psi} = \partial_\mu\bar{\psi} - \bar{\psi}\Gamma_\mu\nn
	\Gamma_\mu &=& \frac{1}{2} w_{\mu\alpha\beta}\sigma^{\alpha\beta}~~\mbox{and}~~ \sigma^{\alpha\beta}=\frac{1}{4}[\gamma^\alpha,\gamma^\beta]~, \nn
	w_{\mu\alpha\beta} &=& e^{\lambda}_{~\alpha}( \partial_\mu e_{\lambda \beta} - \Gamma^{\sigma}_{\lambda\mu} e_{\sigma\beta}) 
	=  e^{\lambda}_{~\alpha}(D_\mu e_{\lambda\beta})~,\nonumber
	\eea
$w_{\mu\alpha\beta}$ being the spin connection.

\subsection{Gauge invariance and degrees of freedom}

The 4-dimensional diffeomorphisms are explicitly broken for both gravity models for $\lambda \neq 1$, $\alpha \neq 0$, 
$F_i\neq0 $ and $S_i\neq 0$. Instead, these models are invariant under 3-dimensional diffeomorphisms 
\bea \label{diffeo}
\delta t&=&f(t)\\
\delta x^{i}&=&\xi^i(t,x)\nn
\delta g_{ij}&=&\partial_{i}\xi_j+\partial_{j}\xi_i+\xi^{k}\partial_{k}g_{ij}+f\dot{g}_{ij} \nn
\delta N_{i}&=&\partial_{i}\xi^kN_k+\xi^k\partial_kN_{i}+\dot{\xi^{j}}g_{ij}+\dot{f}N_i+f \dot{N}_{i} \nn
\delta N&=&\xi^k\partial_kN+\dot{f}N+f \dot{N}~.\nonumber
\eea
Because of 4-dimensional diffeomorphism breaking, a third physical dof is present in both models, and a way of counting the number of dof 
of the theory is given in \cite{Henneaux}, where the Authors use a Hamiltonian description of gauge theories.
They show that the number of primary constraints to take into account is the number of gauge functions {\it plus} 
the number of their time derivatives, in the situation where these gauge functions depend on space and time.
This is because gauge functions and their time derivatives must be considered independently, when defining a boundary condition for the 
evolution of gauge fields.
In our case, we have 10 independent metric components ($N,~N_i,~g_{ij}$) and we see from the gauge transformations (\ref{diffeo}) that the functions 
$\xi^i$ count twice since they appear with their time derivative, while $f$ counts once only because it depends on $t$ only.
The total number of dof is therefore $10-(2\times3+1) = 3$. 

In order to study the one-loop quantum corrections to the matter sectors, we expand the metric $g_{\mu\nu}$ and, consequently, $e^\mu_{~\alpha}$ around a flat background:
\bea 
g_{\mu\nu} &=& \eta_{\mu\nu} + h_{\mu\nu}\\
g^{\mu\nu} &=& \eta^{\mu\nu} -h^{\mu\nu}+h^{\mu\lambda}h_{\lambda}^\nu+\cdots~,\nonumber\\
e^{~\alpha}_{\mu} &=& \delta^{\alpha}_{\mu} + \frac{1}{2} h^{\alpha}_{\mu} - \frac{1}{8} h_{\mu \lambda } h^{\lambda 
\alpha} + \cdots\nonumber\\
e^{\mu}_{~\alpha} &=& \delta^{\mu}_{\alpha} - \frac{1}{2} h^{\mu}_{\alpha} + \frac{3}{8} h^{\mu \lambda } h_{\lambda \alpha} + \cdots ~,\nonumber
\eea
where dots represent higher orders in fluctuations.
Using the following relations
\bea 
g_{\mu\nu} &=& e_{\mu}^{~\alpha} e_{\nu}^{~\beta} \eta_{\alpha\beta}~,\\
e_{\mu}^{~\alpha}e^{\mu}_{~\beta}&=&\eta^{\alpha}_{\beta}~,~~e_{\nu}^{~\alpha}e^{\mu}_{~\alpha}=g^{\mu}_{\nu}~,\nonumber\\
g_{\mu\nu} &=& -N^2 d^2 t + g_{ij}(dx^i+N^idt)(dx^j+N^jdt)~,\nonumber
\eea 
we can then express the different ADM components in terms of their fluctuations
\bea
N &=& 1+n\\
N_i &=& n_i \nonumber\\
g_{ij} &=& \delta_{ij} + h_{ij}~.\nonumber
\eea
The fluctuations $n_i$ and $h_{ij}$ can additionally be decomposed into their different spin components as:
\bea
n_i &=& n_i^T + \partial_i \rho~,\label{nidec}\\
h_{ij} &=& H_{ij}+ (\partial_i W_j+ \partial_j W_i ) + \left(\partial_{i}\partial_{j}-\frac{\delta_{ij}}{3}\partial^2\right)B +\frac{\delta_{ij}}{3}h~,\label{hdecomp}
\eea
where $H_{ij}$ is a transverse-traceless tensor, $n_i^T$ and $W_i$ are transverse vectors and $B$, $h$ and $\rho$ are scalar fields, $h$ being the trace of $h_{ij}$. 
Nevertheless, making use of the gauge freedom shown in eq.(\ref{diffeo}), a natural gauge choice is to set $W_i$ and scalar $B$ to zero. Consequently, eq.(\ref{hdecomp}) becomes
\be\label{hdec} 
h_{ij}=H_{ij} +\frac{\delta_{ij}}{3} h~,
\ee 
where $H_{ij}$ and $h$ are the 3 physical degrees of freedom present in the theory, while $n$, $n_i^T$ and $\rho$ are auxiliary fields.

\subsection{Expanding the actions }

Since we are interested in one-loop corrections, it is enough to expand the actions up to quadratic order in the ADM field fluctuations.
The flat space metric is $\delta_{ij}$, such that 
all the spatial indices are lowered, {\it i.e.} $h^{ij} \to h_{ij} $. We have then
\bea
\sqrt{g} &=& 1 + \frac{1}{2} h + \frac{1}{8} (h^2 - 2h_{ij}h_{ij})+\cdots\\
\Gamma_k &=& -\frac{\sigma_{ij}}{2} \left[  \partial_i h_{kj} - \frac{1}{2}\left(h_{il}\partial_l h_{jk} + h_{lj}\partial_i h_{kl} 
- \frac{1}{2}h_{il}\partial_k h_{jl}\right)\right]+\cdots~,\nonumber
\eea
where $h=h_{ii}$, and dots represent higher orders in fluctuations which can be omitted for the one-loop calculation.

\subsubsection{Matter sector}

We explain here the construction of the relevant matter actions for scalars and fermions, and then describe the ansatz taken for these external fields. 

$\bullet$ Expanding the scalar action (\ref{Ss1}) up to quadratic order in the graviton field fluctuations, the scalar action becomes
\bea\label{Ss2}
S_s^{(2)} &=& -\int dt d^3 x~\left\{\left[1+n+\frac{h}{2}+\frac{hn}{2} +\frac{1}{8}(h^2-2h^2_{ij})\right](-\dot{\phi} \dot{\phi}^\star 
+\partial_k \phi \partial_k \phi^\star)\right.\\
&&+\left. 2 n \dot{\phi}\dot{\phi}^\star+2n_i \dot{\phi}\partial_i \phi^\star-h_{ij}\partial_i\phi\partial_j\phi^\star+2\left(\frac{h n_i}{2}-n n_i 
- n_j h_{ij}\right)\dot{\phi} \partial_i\phi^\star\right.\nonumber\\
&&+\left. (nh-n^2)\dot{\phi} \dot{\phi}^\star
+\left(h_{il}h_{lj}-n_i n_j -nh_{ij}-\frac{h h_{ij} }{2}\right)\partial_i\phi\partial_j\phi^\star\right\}~,\nonumber
\eea
where the first line can only generate Lorentz-symmetric contributions and therefore will be omitted. Moreover, as pointed out in \cite{Pospelov},
quadratic terms in the metric fluctuations will only contribute to our results when the graviton fields are contracted among themselves.
Therefore, for any tensor $T_{ij}$ quadratic in the graviton field, because of rotational invariance in space, one can use the following simplification:
 $T_{ij}\partial_i\phi\partial_j\phi^\star\to (T_{ii}/3)  \partial_k\phi\partial_k\phi^\star$. In addition, linear terms in the metric perturbations can 
 also be omitted, as they lead to quartic matter self interactions after completing the square (we are interested in corrections to the kinetic terms only). Finally, terms of the form $hn_i, nn_i$ or $n_jh_{ij}$ do not contribute since they have to be contracted
with another vector metric fluctuations, leading to terms which are at least cubic in fluctuations. 
The relevant part of the action, containing only terms which generate Lorentz-violating contributions, is then 
\be\label{Ss3}
S_s^{(2)} = -\int dt d^3 x~\left\{ (nh-n^2)\dot{\phi}\dot{\phi}^\star + \frac{1}{3}\left(h_{ij}^2-n_i^2 -nh-\frac{h^2}{2}\right)\partial_k\phi\partial_k\phi^\star\right\}~.
\ee
To simplify even more the expression above, we can write
\be
(nh-n^2)\dot{\phi}\dot{\phi}^\star =- (nh-n^2)\partial_\mu\phi\partial^\mu\phi^\star+(nh-n^2)\partial_k \phi \partial_k \phi^\star~,
\ee
and since the first term on the right-hand side of the expression above is Lorentz-symmetric, we only need to take into account the second one, then~(\ref{Ss3}) becomes:
\be\label{Ss4}
\tilde S_s^{(2)} = -\frac{1}{3}\int dt d^3 x~\left[ h_{ij}^2-n_i^2- 3n^2+2nh-\frac{h^2}{2}\right]\partial_k\phi\partial_k\phi^\star~.
\ee

$\bullet$ For the fermion sector, we find
\bea\label{Sf2'}
S_f^{(2)} &=& -\frac{i}{2}\int dt d^3 x \left\{\left[1+n+\frac{h}{2}+\frac{hn}{2} +\frac{1}{8}(h^2-2h^2_{ij})\right]\bar{\psi} \gamma^{\mu} 
\overleftrightarrow{\partial_\mu}\psi\right.\\
&&+\left. \left(n\delta_{ij} -\frac{1}{2}h_{ij}\right)\left( \bar{\psi}\gamma_i \overleftrightarrow{\partial_j}\psi+\bar{\psi}\{\gamma_i,\Gamma_j^{(1)}\}\psi\right) 
+\bar{\psi} \{\gamma^\mu,\Gamma_\mu^{(1)}\}\psi\right.\nonumber\\    
&&+\left.\frac{1}{2}n_i\left[\bar{\psi}\left(\gamma_i\overleftrightarrow{\partial_0}-\gamma^0\overleftrightarrow{\partial_i}
+\{\gamma_i,\Gamma_0^{(1)}\}-\{\gamma^0,\Gamma_i^{(1)}\} \right) \psi  \right]
+\bar{\psi}\{\gamma^\mu,\Gamma_\mu^{(2)}\}\psi \right.\nonumber\\
&&+\left.\left(\frac{h n_i}{4}-\frac{3}{8} n_j h_{ij} - \frac{n n_i}{4}\right)\bar{\psi} \left(\gamma_i \overleftrightarrow{\partial_0}-\gamma^0\overleftrightarrow{\partial_i} \right)\psi 
+\frac{h}{2} \bar{\psi}\{\gamma^\mu,\Gamma_\mu^{(1)}\}\psi+  \right.\nonumber\\
&&+\left. \frac{1}{8}n_i^2\bar{\psi} \gamma^0\overleftrightarrow{\partial_0}\psi +\left(\frac{3}{8}h_{il}h_{lj}-\frac{h h_{ij}}{4}
+\frac{n(h\delta_{ij}-h_{ij})}{2}-\frac{3}{8}n_in_j\right)\bar{\psi} 
\gamma_i \overleftrightarrow{\partial_j}\psi  \right\}~.\nonumber
\eea
As in the scalar case, the terms on the first line would only contribute to Lorentz-symmetric corrections, hence such terms will be omitted. 
For the same reasons explained above, linear terms in the graviton fields as well as terms proportional to 
$\bar{\psi}\gamma_i\overleftrightarrow{\partial_0}\psi$ or  $\bar{\psi}\gamma_0\overleftrightarrow{\partial_i}\psi$ 
do not contribute to our calculation and therefore will also be omitted from now on. Terms containing the anticommutator of 
a $\gamma$ matrix and the spin connection will not 
contribute as well, since they involve the contraction of a symmetric tensor with an antisymmetric one. 
Finally, as discussed above for the scalar action, when $T_{ij}$ is a given quadratic term in graviton, 
we make the replacement  
$T_{ij}\bar{\psi}\gamma_i\overleftrightarrow{\partial_j}\psi \to  (T_{ii}/3)(\bar{\psi}\gamma_k\overleftrightarrow{\partial_k}\psi)$. 
Thus, the Lorentz-violating fermion action we consider becomes
\bea\label{Sf3}
S_f^{(2)} &=& -\frac{i}{2}\int dt d^3 x \left\{\frac{1}{8}n_i^2\bar{\psi} \gamma^0\overleftrightarrow{\partial_0}\psi 
+\frac{1}{3}\left(\frac{3}{8}h_{ij}^2-\frac{h^2}{4}+n h
-\frac{3}{8}n_i^2\right)\bar{\psi} \gamma_k \overleftrightarrow{\partial_k}\psi\right\}~,
\eea
which, after removing Lorentz-symmetric terms, reduces to
\be\label{Sf4}
\tilde S_f^{(2)} =-\frac{i}{4}\int dt d^3 x \left[\frac{1}{4}h_{ij}^2-\frac{h^2}{6}-\frac{1}{2}n_i^2+\frac{2}{3}n h\right]\bar{\psi} \gamma_k \overleftrightarrow{\partial_k}\psi~.
\ee

$\bullet$ After reducing the matter action to their simplest form (\ref{Ss4}) and (\ref{Sf4}), we then consider a plane-wave as an ansatz for the external fields 
\bea\label{ansatz}
\phi(x)&=& \phi_0\exp(-ip^\mu x_\mu)\nn
\psi(x)&=& \psi_0\exp(-iq^\mu x_\mu)~, 
\eea
so that the 
quantities $\partial_k \phi\partial_k \phi^\star$ and $\bar{\psi}i\gamma_k \overleftrightarrow{\partial_k}\psi$ become constants and, therefore, 
will be respectively replaced by $(\vec p^2\phi_0^2)$ and $(-2i)[\bar{\psi}_0(\vec{\gamma}\cdot\vec q)\psi_0]$. With these classical background matter 
field configurations and using the field decompositions (\ref{nidec}) and (\ref{hdec}), the scalar~(\ref{Ss4}) and fermion~(\ref{Sf4}) actions are respectively
\bea\label{expm}
\tilde S_s^{(2)} &=& -\frac{1}{3}\int dt d^3 x~\left[H_{ij}^2-\frac{h^2}{6}-(n_i^T)^2+\rho\partial^2\rho-3n^2+2nh \right](\vec p^2\phi_0^2)~,\\
\tilde S_f^{(2)} &=& -\frac{1}{2}\int dt d^3 x~\left[\frac{1}{4}H_{ij}^2-\frac{h^2}{12}-\frac{(n_i^T)^2}{2}+\frac{1}{2}\rho \partial^2\rho+
\frac{2}{3}nh\right][\bar{\psi}_0(\vec{\gamma}\cdot\vec q)\psi_0]~.\nonumber
\eea

\subsubsection{Gravity actions} 

For the gravity actions, we expand~(\ref{SG}) and~(\ref{SHL}) up to quadratic order in the metric fluctuations and make use of the metric decompositions~(\ref{nidec}) and~(\ref{hdec}) to obtain 
\bea\label{expg}
S_G^{(2)} &=& M_P^2 \int dt d^3 x \left[ \frac{1}{4}H_{ij}(\partial^2-\partial_t^2)H_{ij} -\frac{1}{2}n_i^T \partial^2 n_i^T-(\lambda-1)\rho (\partial^2)^2\rho\right.\\
&&+\left.\frac{(3\lambda-1)}{12}h \partial_t^2h -\frac{1}{18}h \partial^2h-\alpha n \partial^2 n+\frac{(3\lambda-1)}{3}\rho \partial^2 \dot{h} - \frac{2}{3}n \partial^2 h\right]~,\nonumber
\eea
and
\bea\label{exphl}
S_{HL}^{(2)} &=& M_P^2 \int dt d^3x \left\{ \frac{1}{4} H_{ij} \left[- \partial_t^2 +\partial^2 + F_1 (\partial^2)^2 - S_1 (\partial^2)^3 \right] H_{ij}-\frac{1}{2} n_i^T \partial^2 n_i^T\right. \\
&& + \left. \frac{1}{18} h \left[\frac{3(3\lambda-1)}{2}\partial_t^2-\partial^2 +(3F_1+8F_2)(\partial^2)^2-(3S_1+8S_2)(\partial^2)^3  \right] h\right.\nonumber\\
&& - \left.  n[\alpha\partial^2+F_4 (\partial^2)^2+S_4 (\partial^2)^3] n - (\lambda-1)\rho (\partial^2)^2 \rho +\frac{(3\lambda-1)}{3} \rho \partial^2 \dot{h}\right. \nonumber\\
&& - \left. \frac{2}{3} h [\partial^2+F_3 (\partial^2)^2 +S_3 (\partial^2)^3] n\right\} \nonumber~.
\eea 
Finally, because ghosts do not couple to the matter sector at tree level, one needs to consider 
at least two-loop corrections in order to have a non-vanishing contribution coming from interactions with ghost fields.
Therefore, in the present work, ghosts can be omitted.

\section{One-loop matter effective actions}

We integrate here the metric fluctuations in order to obtain the effective matter kinetic terms. For both gravity models, we impose the Hamiltonian and momentum constraints 
in the path integral, which consists in keeping auxiliary fields on-shell. This approach is used in \cite{Padilla} and implies that no conformal instability arises in our calculations
\cite{Gibbons}. It is known in perturbative quantum gravity that, when introducing an irreducible decomposition for the metric, some of the components have a propagator with 
the wrong sign, leading to a potential problem in defining the partition function. This unstable mode can be traced down to a conformal factor, but with our gauge choice 
it would arise from the integration of the auxiliary fields in the metric decomposition, which does not happen here.

This problem of conformal instability can be understood as an artifact arising from perturbative expansion \cite{Mazur}, but can also be avoided by making an analytical continuation 
of the metric components to imaginary values, simultaneously with the Wick rotation \cite{'tHooft}. 
We note here another non-trivial connection between the Wick rotation and quantum gravity, in the context of time-dependent bosonic strings.
As shown in \cite{AEM} for a specific string configuration, which satisfies conformal invariance non-perturbatively in $\alpha'$, 
a Wick rotation in target space implies a phase factor for the overall string partition function. 
This phase becomes real for specific space-time dimensions only, and 4 appears to be the lowest dimension for which the string partition function remains real, after the Wick rotation.

\subsection{Model I: modified Einstein-Hilbert gravity}

We study here the model (\ref{SG}), for which one-loop quantum corrections are quadratically divergent, and will therefore be regularised with a cut off.
After expansion of the matter and gravity sectors in 
terms of the metric fluctuations, the actions that we are interested in here are (\ref{expm}) and (\ref{expg}).

\subsubsection{Constraints}

Fluctuations in the shift vector $(n_i^T,\rho)$ and the lapse function $(n)$ are auxiliary fields and are therefore not propagating, such that one can use their equations of motion
(the momentum and Hamiltonian constraints) that we substitute back into the action.

Varying the actions with respect to $n$ gives us the following constraint
\be\label{n1constr}
[-2\alpha M_P^2\partial^2 +2(\vec{p}^2\phi_0^2) ] n = \frac{2}{3}\left[M_P^2 \partial^2+ (\vec{p}^2\phi_0^2) +\frac{1}{2} [\bar{\psi}_0(\vec{\gamma}\cdot\vec q)\psi_0]\right]h~,
\ee 
whereas variations with respect to the scalar and transverse parts of the shift vector lead to
\bea  
\left[2M_P^2(1-\lambda)\partial^2 -\frac{2}{3}\left((\vec{p}^2\phi_0^2) + \frac{3}{4}[\bar{\psi}_0(\vec{\gamma}\cdot\vec q)\psi_0]\right)\right]\partial^2 \rho 
&=& -\frac{M_P^2(3\lambda-1)}{3}\partial^2 \dot{h}~,\label{rconstr}\\
\left[-M_P^2\partial^2+\frac{2}{3}\left((\vec{p}^2\phi_0^2) + \frac{3}{4}[\bar{\psi}_0(\vec{\gamma}\cdot\vec q)\psi_0]\right)\right]n_i^T &=& 0~.\label{nTconstr}
\eea  
When the last constraint is put back into the actions, all contributions coming from $n_i^T$ disappear and from now on such actions will only depend on the tensor
and scalar components of the metric. On the other hand, since the auxiliary scalar fields $n$ and $\rho$ appear mixed with the scalar graviton $h$, before 
substituting the constraints~(\ref{n1constr}) and (\ref{rconstr}) back into the actions, we can expand them in terms of the matter contributions to find
\bea\label{nrconstr}  
n&=& -\frac{1}{3\alpha}\left[1+\frac{(\vec{p}^2\phi_0^2)}{M_P^2}\left(\frac{\alpha+1}{\alpha}\right)(\partial^2)^{-1} 
+ \frac{[\bar{\psi}_0(\vec{\gamma}\cdot\vec q)\psi_0]}{2M_P^2}(\partial^2)^{-1}\right]h+\cdots~,\label{nconst1}\\
\rho &=& \frac{(3\lambda-1)}{6(\lambda-1)}\left[1-\frac{(\vec{p}^2\phi_0^2) 
+ \frac{3}{4}[\bar{\psi}_0(\vec{\gamma}\cdot\vec q)\psi_0]}{3(\lambda-1)M_P^2}(\partial^2)^{-1}\right](\partial^2)^{-1}\dot{h}+\cdots,\label{rconst}
\eea  
where dots represent higher-order terms in the matter fields.
The other equations of motion fed back into the actions lead to
\bea\label{SmI}
S^{(2)}_I &=&\int dt d^3x \left\{ \frac{1}{2} H_{ij} \left[ \frac{M_P^2}{2} (\partial^2- \partial_t^2) -\frac{2}{3}(\vec p^2 \phi_0^2)
- \frac{1}{4}[\bar{\psi}_0(\vec{\gamma}\cdot\vec q)\psi_0]\right] H_{ij}\right.\\
&& + \left. \frac{1}{2} h \left[ \frac{M_P^2}{9}\left(-X\partial_t^2+\left(\frac{2-\alpha}{\alpha}\right)\partial^2\right)
+\frac{(\vec{p}^2\phi_0^2)}{9}\left(\frac{\alpha^2+4\alpha+2}{\alpha^2}+\frac{X^2}{6}\frac{\partial_t^2}{\partial^2}\right)\right.\right.\nonumber\\
&& +\left.\left. \frac{[\bar{\psi}_0(\vec{\gamma}\cdot\vec q)\psi_0]}{9}\left(\frac{3\alpha+8}{4\alpha}+\frac{X^2}{8}\frac{\partial^2_t}{\partial^2}\right)\right] h\right\} \nonumber~,
\eea 
where 
\be\label{X} 
X=\frac{3\lambda-1}{\lambda-1}~.
\ee
For a consistent propagation of the scalar graviton $h$, one needs $X>0$, such that the allowed values for $\lambda$ are
\be\label{allowedlambda}
\lambda<1/3~~\mbox{or}~~\lambda>1~.
\ee
Finally, from the action (\ref{SmI}) we find the following dispersion relation for $h$
\be\label{hdrIR}
\omega^2 = \left(\frac{\lambda -1}{3\lambda-1}\right) \left(\frac{2-\alpha}{\alpha}\right) \vec{k}^2~,
\ee 
which shows a consistent propagation for $0<\alpha<2$ for the allowed values of $\lambda$ given in eq.(\ref{allowedlambda}).

\subsubsection{Loop integration}

We give in the Appendix the details of the integration over graviton.\\ 

\nin{\bf Spin-2 component}\\
The integration over the spin-2 component gives 
\be
\exp\left\{-\frac{1}{M_P^2}\left[\frac{4}{3}(\vec p^2 \phi_0^2) + \frac{1}{2}[\bar{\psi}_0(\vec{\gamma}\cdot\vec q)\psi_0]\right]
\frac{\delta(0)}{2(2\pi)^2} \Lambda^2+\cdots\right\}~,\nonumber
\ee 
where $\delta(0)$ is the space-time volume and dots represent either field-independent terms or higher orders in $(\vec p^2 \phi_0^2)$ and $[\bar{\psi}_0(\vec{\gamma}\cdot\vec q)\psi_0]$.

\vspace{0.5cm}

\nin{\bf Spin-0 component}\\
The integration over the spin-0 component gives 
\bea 
&&\exp\left\{\frac{1}{M_P^2}\left[\frac{\alpha^2+4\alpha+2}{2\alpha^2}(\vec p^2\phi_0^2)+\frac{3\alpha+8}{8\alpha}
[\bar{\psi_0}(\vec{\gamma}\cdot \vec q)\psi_0]\right.\right.\nn
&&~~~~~~~~~~~~~~~~~~\left.\left.-\frac{X^2}{4}\left(\frac{(\vec p^2\phi_0^2)}{3}+\frac{[\bar{\psi_0}(\vec{\gamma}\cdot \vec q)\psi_0]}{4}\right)\right]
Y\frac{\delta(0)\Lambda^2}{2(2\pi)^2} +\cdots \right\}~.
\eea
where
\be
Y = \sqrt{ \frac{\alpha(\lambda-1)}{(2-\alpha)(3\lambda-1)} }~.
\ee

\subsubsection{Total Lorentz-violating contributions}

Considering the results obtained in the previous sections, we add here all relevant contributions and present the total Lorentz-violating corrections for both scalar and fermion fields.
We also note that
\be\label{dpf}
(\vec p^2 \phi_0^2) = \frac{1}{\delta(0)}\int dt d^3x~\partial_k \phi \partial_k \phi^\star~~~\mbox{and}~~~[\bar{\psi}_0(\vec{\gamma}\cdot\vec q)\psi_0] 
= \frac{1}{\delta(0)}\int dt d^3x~\bar{\psi}i\gamma_k \partial_k\psi~.
\ee 

Considering the results (\ref{resHI}) and (\ref{reshI}), the total contributions to the matter fields are given by the following Lagrangians 
\be\label{sresI}  
\frac{1}{2(2\pi)^2}\frac{\Lambda^2}{M_P^2}\left[\frac{4}{3}+ \frac{Y}{2}\left(\frac{X^2}{6}-\frac{\alpha^2+4\alpha+2}{\alpha^2}\right)\right]
\partial_k \phi\partial_k \phi^\star~
\ee
in the scalar case, and 
\be\label{fresI}  
\frac{1}{2(2\pi)^2}\frac{\Lambda^2}{M_P^2}\left[\frac{1}{2}-\frac{Y}{8}\left(\frac{3\alpha+8}{\alpha}-\frac{X^2}{2}\right)\right]
\bar{\psi}i\gamma_k\partial_k\psi~,
\ee
in the fermion case,
where $X$ and $Y$ are defined in eq.(\ref{X}) and eq.(\ref{Y}) respectively.

\subsection{Model II: non-projectable HL gravity}

We turn here to the non-projectable version of Horava-Lifshitz gravity~(\ref{SHL}), for which the relevant actions are given by eqs.(\ref{expm}) and (\ref{exphl}). 
In the absence of matter, one-loop quantum corrections for this model are logarithmically divergent, but in the presence of dynamical matter, quadratic divergences arise 
from the coupling gravity-bosonic matter \cite{Pospelov,Padilla}.
In our case, matter cannot induce further divergences compared to the single gravity case, since it is classical and thus does not involve any new loop momentum.
The use of Hamiltonian and momentum constraints generate an artificial quartic divergence though, due to the introduction of additional space derivatives in the 
decompositions (\ref{nidec},\ref{hdecomp}) of the graviton: as can be seen from the action (\ref{SmII}) below, the presence of matter is then accompanied with the derivative operator
$\partial_t^2/(\vec\partial)^2$, arising from the coupling between $\rho$ and $h$ in the gravity sector. 
This divergence is thus a gauge artifact, on which we will come back in section 4, and that we disregard in our calculations.

As a consequence, 
we use dimensional regularisation, where the naively power-law divergent integrals actually vanish \cite{Leibbrandt}.
We note that the vanishing or finiteness of a regularised integral which otherwise would naively be divergent is explained pedagogically in \cite{Weinzierl}: in the regularised integral, 
divergences associated to different regions of the domain of integration cancel each other, such that the integral is finite when the regulator is removed.

\subsubsection{Constraints} 

The constraints (\ref{rconstr}) and (\ref{nTconstr}) obtained from the variation of the action with respect to the shift vector components ($n_i^T,\rho$) are the same as in the previous 
modified gravity model, since the  dimension 4 and dimension 6 operators which are added do not depend on these components.
The additional contributions to the lapse function fluctuations $n$ lead to the new constraint
\be 
[-2M_P^2 \mathcal{D}_2 + 2 (\vec{p}^2\phi_0^2)]n =\frac{2}{3}\left[M_P^2 \mathcal{D}_1 + (\vec{p}^2\phi_0^2) + \frac{1}{2}[\bar{\psi}_0(\vec{\gamma}\cdot\vec q)\psi_0] \right] h ~,
\ee 
which can be written as
\be\label{nconst}
n = -\frac{1}{3}\left[\frac{\mathcal{D}_1}{\mathcal{D}_2} + \frac{(\vec{p}^2 \phi_0^2)}{M_P^2}\frac{1}{\mathcal{D}_2}\left(1+\frac{\mathcal{D}_1}{\mathcal{D}_2}\right) 
+ \frac{1}{2M_P^2}[\bar{\psi}_0(\vec{\gamma}\cdot\vec q)\psi_0]\frac{1}{\mathcal{D}_2} \right]h+\cdots~,
\ee 
where dots represent higher-order terms in $[\bar{\psi}_0(\vec{\gamma}\cdot\vec q)\psi_0]$ and $(\vec{p}^2\phi_0^2)$, and
\bea 
\mathcal{D}_1 &=& [\partial^2+ F_3 (\partial^2)^2+S_3 (\partial^2)^3]~,\\
\mathcal{D}_2 &=& [\alpha\partial^2+ F_4 (\partial^2)^2+S_4 (\partial^2)^3]~.\nonumber
\eea 
Using the constraints (\ref{rconstr}), (\ref{nTconstr}) and (\ref{nconst}) to rewrite the original actions~(\ref{expm}) and~(\ref{exphl}), we arrive at the following action, 
which only depends on the physical metric fluctuations $H_{ij}$ and $h$,
\bea\label{SmII} 
S^{(2)}_{II} &=&\int dt d^3x \left\{ \frac{1}{2} H_{ij} \left[ \frac{M_P^2}{2} (- \partial_t^2 +\partial^2 + F_1 (\partial^2)^2 - S_1 (\partial^2)^3 ) -\frac{2}{3}(\vec p^2 \phi_0^2)
- \frac{1}{4}[\bar{\psi}_0(\vec{\gamma}\cdot\vec q)\psi_0]\right] H_{ij}\right.\nonumber \\
&& + \left. \frac{1}{2} h \left[ \frac{M_P^2}{9}\left(-X\partial_t^2-\partial^2+(3F_1+8F_2)(\partial^2)^2-(3S_1+8S_2)(\partial^2)^3
+2\left(\frac{\mathcal{D}_1^2}{\mathcal{D}_2}\right)\right)\right.\right.\nonumber\\
&&+\left.\left. \frac{(\vec{p}^2\phi_0^2)}{9}\left(1+4\left(\frac{\mathcal{D}_1}{\mathcal{D}_2}\right)+2\left(\frac{\mathcal{D}_1}{\mathcal{D}_2}\right)^2
+\frac{X^2}{6}\frac{\partial_t^2}{\partial^2}\right)\right.\right.\nonumber\\
&& +\left.\left. \frac{[\bar{\psi}_0(\vec{\gamma}\cdot\vec q)\psi_0]}{9}\left(\frac{3}{4}+2\left(\frac{\mathcal{D}_1}{\mathcal{D}_2}\right)
+\frac{X^2}{8}\frac{\partial^2_t}{\partial^2}\right)\right] h\right\}~.
\eea

\subsubsection{Loop integration}

We give in the Appendix the details of the integration over graviton, which is done using 
dimensional regularisation, with $d=3-\epsilon$.\\

\nin{\bf Spin-2 component}\\
The integration over the spin-2 component gives
\be 
\exp\left\{-\frac{1}{M_P^2}\left[\frac{4}{3}(\vec p^2 \phi_0^2) + \frac{1}{2}[\bar{\psi}_0(\vec{\gamma}\cdot\vec q)\psi_0]\right]
\frac{\delta(0)}{(2\pi)^2\sqrt{|S_1|}}\frac{\mu^\epsilon}{\epsilon}+\cdots\right\}~.
\ee 

\vspace{0.5cm}

\nin{\bf Spin-0 component}\\
The integration over the spin-0 component gives
\bea 
&&\exp\left\{\frac{1}{M_P^2}\left[\frac{1}{2\sqrt{C_6^{(0)}}}\left((\vec p^2 \phi_0^2)+\frac{3}{4}[\bar{\psi}_0(\vec{\gamma}\cdot\vec q)\psi_0]\right)
+\frac{(\vec p^2 \phi_0^2)}{\sqrt{C_6^{(2)}}}\right.\right.\\
&&\left.\left.~~~~~~~~~~~~~~~~~~~~~~~~~~+\frac{1}{\sqrt{C_6^{(1)}}}\left(2(\vec p^2 \phi_0^2)+[\bar{\psi}_0(\vec{\gamma}\cdot\vec q)\psi_0]\right)\right]
\frac{\delta(0)}{(2\pi)^2\sqrt{X}}\frac{\mu^\epsilon}{\epsilon}+\cdots\right\}\nonumber~,
\eea
where $C_6^{(n)}$, with $n=0,1,2$, are given in the Appendix.

\subsubsection{Total Lorentz-violating contributions}

In the Appendix appears the following integral
\be\label{intd}
\mathcal{I}\left(\Delta\right)=\int \frac{d^d k}{(2\pi)^d} \frac{1}{\vec{k}^2\sqrt{\vec{k}^2+ \Delta }}~,
\ee
which, with dimensional regularisation ($d=3-\epsilon$), gives 
\be\label{Isol}  
\mathcal{I}(\Delta) = \frac{1}{2 \pi^2} \frac{\mu^\epsilon}{\epsilon}+\mathcal{O}(\epsilon)~.
\ee 
In the limit $\epsilon\to0$, we obtain then
\be 
\mu \frac{\partial}{\partial \mu} \mathcal{I}(\Delta)= \frac{1}{2 \pi^2}~,
\ee 
such that 
\be\label{Isol2}
\mathcal{I}(\Delta) = \frac{1}{2 \pi^2} \ln\left(\frac{\mu}{\mu_0}\right)~,
\ee 
where $\mu_0$ is a mass scale. To choose $\mu_0$ accordingly, we calculate $\mathcal{I}(\Delta)$ using a cut off $\Lambda$ in dimension $d=3$ and find:
\be\label{Ico} 
\mathcal{I}(\Delta) = \frac{1}{2\pi^2}\ln\left(\frac{\Lambda+\sqrt{\Lambda^2 +\Delta}}{\sqrt{\Delta}}\right)~.
\ee 
From the form of $\Delta$ in~(\ref{trH}) and (\ref{3dint}), we note that $\Delta$ can be written as $d M_{HL}^2$, where $d$ represents a 
dimensionless constant of order 1 for each of the different cases. Then, expanding~(\ref{Ico}) for $\Lambda\gg M_{HL}$, we find
\be 
\mathcal{I}(\Delta) = \frac{1}{2\pi^2} \ln\left(\frac{\Lambda}{M_{HL}}\right) ~,
\ee 
where finite terms were omitted in the expression above.
Comparing eq.(\ref{Isol2}) with the result above, we naturally choose $\mu_0 = M_{HL}$ and $\mu=\Lambda$.

Considering now the results obtained above, we write the total Lorentz-violating contributions for both scalar 
and fermion fields. Using the relations (\ref{dpf}) and assuming~(\ref{Isol2}) with $\mu_0=M_{HL}$, the total corrections for scalar and fermion fields are, respectively,
\bea \label{sfresII}
&&\left[-\frac{1}{2(2\pi)^2}\left(\frac{4}{3}\frac{1}{\sqrt{|s_1|}} - \frac{1}{2}\frac{1}{\sqrt{X c_6^{(0)}}} -\frac{2}{\sqrt{X c_6^{(1)}}}-\frac{1}{\sqrt{X c_6^{(2)}}}  \right) 
\frac{M_{HL}^2}{M_P^2} \ln\left(\frac{M_{HL}^2}{\Lambda^2 }\right)\right] \partial_k \phi \partial_k \phi^\star~,\nonumber\\
&&\left[-\frac{1}{2(2\pi)^2}\left(\frac{1}{2}\frac{1}{\sqrt{|s_1|}} - \frac{3}{8}\frac{1}{\sqrt{X c_6^{(0)}}} -\frac{1}{\sqrt{X c_6^{(1)}}} \right) \frac{M_{HL}^2}{M_P^2} 
\ln\left(\frac{M_{HL}^2}{\Lambda^2 }\right)\right]\bar{\psi} i \gamma_k \partial_k \psi~.
\eea

\subsection{Non-minimal coupling}

The models studied here involve minimally coupled matter, and one could ask what the effect of a non-minimal coupling would have on the effective 
dispersion relations. We show here that these would not affect our results.

Given that we impose all the terms in the action to be at most of mass dimension 6, and that the scalar field is dimensionless for $z=3$ in 
$d=3$ space dimensions, its non-minimal coupling to gravity would be for example of the form
\be
\left[\xi_1 R^{(3)}+\xi_2(R^{(3)})^2+\xi_3(a_i a^i)\right](\phi\phi^\star)~,
\ee
where $\xi_i$ have the appropriate mass dimension for the terms inside the square bracket to be of dimension 6.
Similarly, for the fermion of mass dimension 3/2, we could have
\be
\left[\zeta_1 R^{(3)}+\zeta_2(a_i a^i)\right](\ol\psi\psi)~.
\ee
In both cases, the space derivatives appearing in $R^{(3)}$ and $a_i$ generate space derivatives for matter fields, after integration by parts,
and thus could naively contribute to the effective dispersion relation. But the 
latter is obtained with matter plane wave configurations, such that $\phi\phi^\star$ and $\ol\psi\psi$ are constants, and these configurations would therefore not 
give additional contributions to the dispersion relation. A more general field configuration could be chosen of course, but which would also contribute 
to all the other terms calculated in this article, in such a way that the final dispersion relation would not change: the functional for matter fields obtained after 
integrating gravitons is unique, and the corresponding dispersion relation is obtained by plugging a plane wave solution.

This conclusion is valid at one-loop though: non-minimal coupling would radiatively generate terms which would modify the matter kinetic terms, with an impact
on the dispersion relation at two loops and higher orders.

\subsection{Comments on regularisation}

All the integrals in this article have been calculated by first integrating over frequency and then over momentum.
In most of the integrals, the integration over frequency is finite, and has been performed without any regularisation. We are then left with a 3-dimensional integration
over momentum, which is regularised either with a cut off or with dimensional regularisation.
But there is also the situation where the integration over frequency is divergent, and we discuss here few details for both models.\\

\nin $\bullet$ Model I

As discussed previously, this model involves quadratic divergences and therefore cannot be treated with dimensional regularisation, which sees only logarithmic 
divergences. A typical example where the integration over frequencies is finite is
\be 
\int \frac{d^4k}{(2\pi)^4} \frac{1}{\omega^2+z^2\vec{k}^2} 
=\frac{1}{4\pi^3} \int_0^\Lambda dk\int_{-\infty}^{+\infty} d\omega \frac{\vec{k}^2}{\omega^2+z^2\vec{k}^2}
=\frac{\Lambda^2}{8\pi^2z}~.\\
\ee
where $z$ is a positive dimensionless constant. \\
But we also have integrals for which the integration over frequencies is divergent, and 
these are calculated with the same cut off as the one used for momentum integration in the other integrals. The integral is of the form
\bea
\int \frac{d^4k}{(2\pi)^4} \frac{\omega^2}{\vec{k}^2(\omega^2+z^2\vec{k}^2)}
&=&\frac{1}{4\pi^3} \int_0^{\infty} dk\int_{-\Lambda}^{+\Lambda} d\omega \frac{\omega^2}{\omega^2+z^2\vec{k}^2}\\
&=&\frac{1}{4\pi^3}\int_0^{\infty} dk\int_{-\Lambda}^{+\Lambda} d\omega \left[1-\frac{z^2\vec{k}^2}{\omega^2+z^2\vec{k}^2}\right]\nn
&=&\frac{1}{2\pi^3}\int_0^{\infty} dk\left[\Lambda- zk \arctan\left(\frac{\Lambda}{z k}\right)\right]\nn
&=&\frac{\Lambda^2}{8\pi^2z}\nonumber~.
\eea
We note that the same result is obtained if one performs first the finite integration over momentum, and then uses the cut off for frequency, whereas the commutativity of 
the order of integration is not obvious when the integrals are divergent.\\

\nin $\bullet$ Model II

The second model contains only logarithmic divergences, but artificial quartic divergences appear as a result of the graviton decomposition (\ref{hdecomp})
in terms of auxiliary fields. Since $[\omega]=3$, we regularise the integral over frequency with the cut off $\Lambda^3$
\bea
\int \frac{d^4k}{(2\pi)^4} \frac{\omega^2}{\vec{k}^2(\omega^2+z^6\vec{k}^6)} &=& \frac{1}{4\pi^3}
\int_0^{\infty} dk\int_{-\Lambda^3}^{+\Lambda^3} d\omega \left[1-\frac{z^6\vec{k}^6}{\omega^2+z^6\vec{k}^6}\right]\\
&=&\frac{1}{2\pi^3}\int_0^{\infty} dk \left[\Lambda^3-z^3k^3 \arctan\left(\frac{\Lambda^3}{z^3k^3}\right)  \right]\nn
&=&\frac{\Lambda^{4}}{8\pi^2z}\nonumber~.
\eea
As explained at the beginning of section 3.2, this divergence is artificial and is thus omitted.
The other integrals for this model involve a finite integration over frequency, and a logarithmically divergent momentum integral. The latter is 
regularised with dimensional regularisation, but the same coefficient of the logarithmic divergence would be found with a cut off, as explained in section 3.2.3.
Therefore the Lorentz-symmetry violating terms generated in this model are unique (in the present gauge) and independent of the regularisation.

\section{Analysis}

The Lorentz-symmetry violating effect is measured by the product $v^2$ of group and phase velocities, which should be equal to 1 
in the Lorentz-symmetric case.
As noted in \cite{Pospelov}, the measurable deviation from the Lorentz-symmetric case is the difference $|v_s^2-v_f^2|$, 
which should be typically smaller than
$10^{-20}$, according to the current upper bounds on Lorentz-symmetry violation \cite{bounds}. We note that \cite{Pospelov}
consider dynamical matter fields, instead of classical ones, which allows the cancellation of the above mentioned quartic divergence, as expected  
for a gauge artifact. Indeed, the graviton loop giving rise to this divergence is canceled by the equivalent matter loop, the difference of 
signs coming from the fact that the gravity spin-0 field component leading to the quartic divergence is auxiliary.
In our case though, this cancellation does not occur because the matter loop is not present. This shows the non-trivial fact that a classical
matter background can be consistently taken into account only if one removes this specific gauge artifact by hand. Other gauge artifacts are removed by
taking the difference  $|v_s^2-v_f^2|$, as we do below.

The calculation of the traces obtained in the previous sections leads to the following kinetic terms
\bea
&&-i\bar{\psi} \gamma^0 \partial_t\psi+v_f^{(m)} i\bar{\psi}\gamma_k \partial_k\psi\\
&&-\partial_t\phi\partial_t\phi^\star+(v_s^{(m)})^2\partial_k\phi\partial_k\phi^\star~,\nonumber
\eea
where $v_f^{m} = 1+\delta v_f^{m}$ and $(v_s^{m})^2=1+(\delta v_s^{m})^2$ and $m =I, II$ for the first model (\ref{SG}) and the second model (\ref{SHL}) 
respectively.

\subsection{Model I}

For the first model, $(\delta v_s^{I})^2$ and $\delta v_f^{I}$ are given by eqs.(\ref{sresI}) and (\ref{fresI}), respectively, {\it i.e.}
\bea
(\delta v_s^I)^2 &\simeq& \frac{1}{2(2\pi)^2}\left[\frac{4}{3}
+\frac{Y}{2}\left(\frac{X^2}{6}-\frac{\alpha^2+4\alpha+2}{\alpha^2}\right)\right]\frac{\Lambda^2}{M_P^2}~,\\
(\delta v_f^I)^2&\simeq& 2 \delta v_f^I \simeq  \frac{1}{2(2\pi)^2}\left[1-\frac{Y}{4}\left(\frac{3\alpha+8}{\alpha}-\frac{X^2}{2}\right)\right]\frac{\Lambda^2}{M_P^2}~.
\eea
Subtracting the fermion contribution from the scalar one, we get the measurable departure from Lorentz symmetry
\bea\label{finalresI}
&&|(\delta v_s^I)^2-(\delta v_f^I)^2| = \frac{1}{2(2\pi)^{2}}\left|F(\lambda,\alpha)\right|\frac{\Lambda^2}{ M_P^2}~, ~~~\mbox{where}\\
&&F(\lambda,\alpha) = -\frac{1}{3}+\frac{1}{4}\sqrt{\frac{\alpha(\lambda-1)}{(2-\alpha)(3\lambda-1)}}\left[\frac{(3\lambda-1)^2}{6(\lambda-1)^2}
-\frac{\alpha^2-4}{\alpha^2}\right].\nonumber
\eea
If one does not impose any specific values for $\lambda$ and $\alpha$, $F(\lambda,\alpha)$ is generically of order $1$, such that 
the current experimental bounds on Lorentz violation are satisfied with the difference (\ref{finalresI}) if the graviton loop momentum is
cut off by $\Lambda\lesssim10^{10}$ GeV. 

Nevertheless, one can always find specific values for $\lambda$ and $\alpha$ such that
the difference (\ref{finalresI}) vanishes. An example is 
\be
F(\lambda_0,\alpha_0)=0 ~~~\mbox{for}~~~\lambda_0 \simeq 0.332~~~ \mbox{and}~~~ \alpha_0 \simeq 1.995~,
\ee
which are both in the range of allowed values for these parameters. This set of values has been chosen here because they are close to the  
boundary values $\lambda_b=1/3$ and $\alpha_b=2$, which suggest that quantum corrections point towards this specific point in parameter space.
It is interesting to note that $\lambda_b=1/3$ is found as an IR fixed point for the Wilsonian renormalisation flows studied in \cite{Dodorico},
and it would be interesting to relate this result to ours.

\subsection{Model II}

For the second model, from the results in~(\ref{sfresII}) we can write $(\delta v_s^{II})^2$ and $\delta v_f^{II}$ as
\bea   
(\delta v_s^{II})^2 &\simeq& -\frac{1}{2(2\pi)^2}\left(\frac{4}{3}\frac{1}{\sqrt{|s_1|}} - \frac{1}{2}\frac{1}{\sqrt{X c_6^{(0)}}} 
-\frac{2}{\sqrt{X c_6^{(1)}}}-\frac{1}{\sqrt{X c_6^{(2)}}}  \right) \frac{M_{HL}^2}{M_P^2} \ln\left(\frac{M_{HL}^2}{\Lambda^2 }\right)~,\nonumber\\
(\delta v_f^{II})^2 &\simeq& 2\delta v_f^{II} \simeq -\frac{1}{2(2\pi)^2}\left(\frac{1}{\sqrt{|s_1|}} - \frac{3}{4}\frac{1}{\sqrt{X c_6^{(0)}}} 
-\frac{2}{\sqrt{X c_6^{(1)}}} \right) \frac{M_{HL}^2}{M_P^2} \ln\left(\frac{M_{HL}^2}{\Lambda^2 }\right)~.
\eea 
Consequently, the physical quantity in which we are interested, {\it i.e.} the difference between these two contributions, is
\bea\label{finalresII} 
|(\delta v_s^{II})^2 - (\delta v_f^{II})^2| \simeq\left|\frac{1}{2(2\pi)^2}\left(\frac{1}{3}\frac{1}{\sqrt{|s_1|}} 
+ \frac{1}{4}\frac{1}{\sqrt{X c_6^{(0)}}} -\frac{1}{\sqrt{X c_6^{(2)}}}  \right) \frac{M_{HL}^2}{M_P^2} \ln\left(\frac{M_{HL}^2}{\Lambda^2 }\right)\right|~.
\eea 
If one assumes that $\Lambda \sim M_P$ and $s_1$, $c_6^{(n)}$ are of order 1, the bounds on Lorentz violation~\cite{bounds} 
are then satisfied for $M_{HL}\lesssim10^{10}$ GeV. 
Note that, by construction of HL gravity, $M_{HL}$ should also be large enough for higher-order space derivatives to be small compared to relativistic terms in the IR.
The fact that quantum corrections imply an {\it upper} bound on $M_{HL}$ is because we impose the UV regime, above $M_{HL}$, to dominate the loop integrals ``early'' enough, 
for quantum corrections not to be too large.

However, similarly to what we find with the first model, 
the Lorentz-violating correction in~(\ref{finalresII}) can also be canceled if the coupling constants are chosen accordingly. For instance, one of 
the many ways in which this cancellation takes place is achieved by setting $s_1$, $c_6^{(n)}$ to 1, and taking $\lambda\approx 1.969$, which is in the allowed regime for this parameter.

\subsection{Conclusion}

Our results lead to the following main points 
\begin{itemize}

\item If one wishes to conclude with generic values for the different parameters, then both models lead to the same order of magnitude $10^{10}$ GeV for the typical scale
above which the predicted Lorentz symmetry violation is too large. Therefore, although the second model has an improved behaviour in terms of UV divergences, the IR
phenomenology is not really improved compared to the model without higher-order space derivatives;

\item If one accepts to fine-tune the different parameters, then there is always a range of values for these parameters, such that the effective maximum speed  
seen by particles is consistent with Special Relativity. Once again, this is valid for both models, and this feature is not a consequence of introducing higher-order space derivatives.

\end{itemize}

Finally, the typical scale $10^{10}$ GeV 
is consistent with other modified gravity model, as in \cite{Pospelov}. The same characteristic scale is also found in \cite{AB2},
where non-relativistic corrections to matter kinetic terms are calculated in the framework of the Covariant HL gravity \cite{covariant}.
We note that this scale also corresponds to the Higgs potential instability \cite{Higgsinst}, which could be
avoided by taking into account curvature effects in the calculation of the Higgs potential \cite{rajantie}. It would therefore be interesting
to look for a stabilising mechanism in the framework of non-relativistic gravity models.

A next step consists in deriving higher-order space derivatives for the fermion kinetic term, from a 
4-dimensional diffeomorphism breaking gravity model, 
in order to generate fermion mass and flavour oscillations dynamically, as was shown in \cite{ALM}.

\section*{Appendix: loop calculations}

\subsection*{Model I}

\nin{\bf Spin-2 component}

Gathering all the terms depending on $H_{ij}$ in the actions~({\ref{expm}}) and~({\ref{expg}}), we obtain 
\bea
\tilde S_I^{(2)}(H_{ij}) = \int dt d^3 x~ \frac{1}{2}H_{ij}\left[ \frac{M_P^2}{2} (\partial^2-\partial_t^2) -\frac{2}{3}(\vec p^2 \phi_0^2)
- \frac{1}{4}[\bar{\psi}_0(\vec{\gamma}\cdot\vec q)\psi_0]\right]H_{ij}~,
\eea
which can be rewritten in terms of the Fourier transform $\tilde H_{ij}(k)$ of $H_{ij}(x)$ after a Wick rotation ($t\to it$, $\omega\to-i\omega$) as
\bea
&&\int \mathcal{D}H_{ij} \exp\left\{i \tilde S_{I}^{(2)}(H_{ij})\right\}\to\int \mathcal{D} \tilde{H}_{ij}\exp\left\{-\int \frac{d^4 k_1d^4 k_2}{(2\pi)^8}
~\frac{1}{2}\tilde{H}_{ij}(k_2) \left[\mathcal{A}^{(I)}_{\tilde H_{ij}} \right]  \tilde{H}_{ij}(k_1)\right\}, \nonumber\\
&& \mbox{where}~~ \mathcal{A}^{(I)}_{\tilde H_{ij}} = \left[ \frac{M_P^2}{2} k_1^2 +\frac{2}{3}(\vec p^2 \phi_0^2)
+ \frac{1}{4}[\bar{\psi}_0(\vec{\gamma}\cdot\vec q)\psi_0]\right]\delta(k_1+k_2)\nonumber
\eea
with $k_1^2=\omega_1^2+\vec{k}_1^2$ in Euclidean space. Then, considering the two components of $\tilde{H}_{ij}$, we can perform the functional integration to find
\bea\label{resHI}
&&\left\{\det \left[\frac{1}{(2\pi)^8} \left(\frac{M_P^2}{2} k_1^2 +\frac{2}{3}(\vec p^2 \phi_0^2)+ 
\frac{1}{4}[\bar{\psi}_0(\vec{\gamma}\cdot\vec q)\psi_0] \right)\delta(k_1+k_2) \right]\right\}^{-1} \\
&=& \exp\left\{-\mbox{Tr} \ln\left[\frac{1}{(2\pi)^8}\left( \frac{M_P^2 k_1^2}{2} +\frac{2}{3}(\vec p^2 \phi_0^2) 
+ \frac{1}{4}[\bar{\psi}_0(\vec{\gamma}\cdot\vec q)\psi_0]\right) \right]\delta(k_1+k_2)\right\}\nonumber\\
&=& \exp\left\{-\frac{1}{M_P^2}\left[\frac{4}{3}(\vec p^2 \phi_0^2) + \frac{1}{2}[\bar{\psi}_0(\vec{\gamma}\cdot\vec q)\psi_0]\right]
\mbox{Tr}\left(\frac{\delta(k_1+k_2)}{k_1^2}\right)+\cdots\right\}~,\nonumber
\eea 
where dots represent field-independent terms or higher orders in $(\vec p^2 \phi_0^2)$ and $[\bar{\psi}_0(\vec{\gamma}\cdot\vec q)\psi_0]$. 
The trace is calculated using a momentum cut off $\Lambda$ and leads to
\be\label{trace}
\mbox{Tr}\left(\frac{\delta(k_1+k_2)}{k_1^2}\right) = \int \frac{d^4 k_1d^4 k_2}{(2\pi)^8}\left(\frac{\delta(k_1+k_2)}{k_1^2}\right)\delta(k_1+k_2)= \frac{\delta(0)}{2(2\pi)^2} \Lambda^2~,
\ee 
where $\delta(0)$ is the space-time volume.

\vspace{1cm}

\nin{\bf Spin-0 component}

Considering the terms which depend on the Fourier transform $\tilde h$ in the action (\ref{SmI}) and performing a Wick rotation, we find
\bea
\int \mathcal{D}h \exp\left\{i S_{I}^{(2)}(h)\right\}\to\int \mathcal{D} \tilde{h}\exp\left\{-\int \frac{d^4 k_1}{(2\pi)^4}\frac{d^4 k_2}{(2\pi)^4}
~\frac{1}{2}\tilde{h}(k_2) \left[\mathcal{A}^{(I)}_{\tilde h}\right]\tilde{h}(k_1)\right\}~,
\eea 
where
\bea\label{AhI} 
\mathcal{A}^{(I)}_{\tilde h} &=& \frac{1}{9} \left\{M_P^2\left(X\omega_1^2+\frac{2-\alpha}{\alpha}\vec{k}_1^2\right)+(\vec p^2 \phi_0^2)
\left[\frac{X^2}{6}\frac{\omega_1^2}{\vec{k}_1^2}-\frac{\alpha^2+4\alpha+2}{\alpha^2}\right]\right.\nonumber\\
&&+\left.[\bar{\psi}_0(\vec{\gamma}\cdot\vec q)\psi_0]\left[\frac{X^2}{8}\frac{\omega_1^2}{\vec{k}_1^2}-\frac{3\alpha+8}{4\alpha}\right]\right\}\delta(k_1+k_2)~.
\eea 
The evaluation of the functional integral leads to
\bea \label{reshI}
&&\exp\left\{\frac{1}{M_P^2}\left[\left(\frac{\alpha^2+4\alpha+2}{2\alpha^2}(\vec p^2\phi_0^2)+\frac{3\alpha+8}{8\alpha}
[\bar{\psi_0}(\vec{\gamma}\cdot \vec q)\psi_0]\right)\mbox{Tr}\left(\frac{\delta(k_1+k_2)}{X\omega_1^2+\alpha^{-1}(2-\alpha)\vec{k}_1^2}\right)\right.\right.\nonumber\\
&&-\left.\left.\frac{X^2}{4}\left(\frac{(\vec p^2\phi_0^2)}{3}+\frac{[\bar{\psi_0}(\vec{\gamma}\cdot \vec q)\psi_0]}{4}\right)
\mbox{Tr}\left(\frac{\delta(k_1+k_2)\omega_1^2}{\vec{k}_1^2 \left(X\omega_1^2+\alpha^{-1}(2-\alpha)\vec{k}_1^2 \right)}\right) \right] \right\}~.
\eea 
Both traces above are evaluated with the common cut off $\Lambda$ for frequencies and wave vectors, and lead to the same result  
\bea 
&&Tr\left[\frac{\delta(k_1+k_2)\omega_1^2}{\vec{k}_1^2 \left(X\omega_1^2+\alpha^{-1}(2-\alpha)\vec{k}_1^2 \right)}\right]
=Tr\left[\frac{\delta(k_1+k_2)}{ \left(X\omega_1^2+\alpha^{-1}(2-\alpha)\vec{k}_1^2 \right)}\right] = Y\frac{\delta(0)\Lambda^2}{2(2\pi)^2}~,
\eea 
where
\be\label{Y}
Y = \sqrt{ \frac{\alpha(\lambda-1)}{(2-\alpha)(3\lambda-1)} }~.
\ee

\subsection*{Model II}

\nin{\bf Spin-2 component}

Comparing the terms which depend on $H_{ij}$ in the expressions~(\ref{SmI}) and (\ref{SmII}), 
it is clear that the only difference is on the appearance of higher-order space derivatives in the propagator of the tensor field in the npHL case. Therefore 
the integration over the spin-2 component of the metric in the present model leads to the same result as in~(\ref{resHI}) when $k_1^2=\omega^2+\vec{k}^2_1$ 
(in Euclidean space) is replaced by $\omega_1^2+\vec{k}_1^2-F_1(\vec{k}_1^2)^2-S_1(\vec{k}_1^2)^3$. Assuming that $F_1$ and $S_1$ are negative constants, we obtain then 
\be\label{resHII} 
\exp\left\{-\frac{1}{M_P^2}\left[\frac{4}{3}(\vec p^2 \phi_0^2) + \frac{1}{2}[\bar{\psi}_0(\vec{\gamma}\cdot\vec q)\psi_0]\right]
\mbox{Tr}\left(\frac{\delta(k_1+k_2)}{\omega_1^2+\vec{k}_1^2+|F_1|(\vec{k}_1^2)^2+|S_1|(\vec{k}_1^2)^3}\right)+\cdots\right\}~.
\ee 
\\
In order to calculate the trace above, we first integrate over the frequencies
\bea \label{trH}
\mbox{Tr}\left(\frac{\delta(k_1+k_2)}{\omega_1^2+\vec{k}_1^2+|F_1|(\vec{k}_1^2)^2+|S_1|(\vec{k}_1^2)^3}\right) &=& \frac{\delta(0)}{2} \int \frac{d^3k}{(2\pi)^3} 
\frac{1}{\sqrt{\vec{k}^2+|F_1| \vec{k}^4 +|S_1|\vec{k}^6}}\nn
&\approx&\frac{\delta(0)}{2\sqrt{|S_1|}} ~\mathcal{I}\left(\frac{|F_1|}{|S_1|}\right)~,
\eea 
where
\be 
 \mathcal{I}\left(\Delta\right)=\int \frac{d^3 k}{(2\pi)^3} \frac{1}{\vec{k}^2\sqrt{\vec{k}^2+ \Delta }}~.
\ee 
Using dimensional regularisation, this integral becomes
\be
\mathcal{I}\left(\Delta\right) = \frac{2 \pi^{d/2}}{\Gamma(d/2)} \frac{\mu^{3-d}}{(2\pi)^d}\int_0^\infty d k \frac{k^{d-3}}{(k^2+\Delta)^{1/2}}~,
\ee 
which is calculated using \cite{Ryder}
\be\label{intGamma} 
\int_0^\infty d k \frac{k^\beta}{(k^2+\Delta)^\alpha} = \frac{\Gamma(\frac{1+\beta}{2})\Gamma(\alpha-\frac{1+\beta}{2})}{2(\Delta)^{\alpha-\frac{1+\beta}{2}}\Gamma(\alpha)}~.
\ee 
Writing $d=3-\epsilon$, we find  then
\be
\mathcal{I}(\Delta) = \frac{1}{2 \pi^2} \frac{\mu^\epsilon}{\epsilon}+\mathcal{O}(\epsilon)~,
\ee 
which finally leads to
\be
\mbox{Tr}\left(\frac{\delta(k_1+k_2)}{\omega_1^2+\vec{k}_1^2+|F_1|(\vec{k}_1^2)^2+|S_1|(\vec{k}_1^2)^3}\right) \approx \frac{\delta(0)}{(2\pi)^2\sqrt{|S_1|}}\frac{\mu^\epsilon}{\epsilon}+\cdots
\ee
where dots represent finite terms.
We note here that the above result does not depend on $|F_1|$, since this coupling constant controls $\vec{k}^4$ in~(\ref{resHII}), which is sub-dominant, and
the UV behaviour is dominated by the terms $|S_1|\vec{k}^6$.

\vspace{0.5cm}

\nin{\bf Spin-0 component}

Gathering all terms which depend on $h$ in the action (\ref{SmII}), we write the equivalent action in terms of the Fourier transform $\tilde{h}$ 
of $h$ and, after performing a Wick rotation, we obtain
\bea\label{hwr}
\int \mathcal{D}h \exp\left\{i S_{II}^{(2)}(h)\right\} = \int \mathcal{D} \tilde{h}\exp\left\{-\int \frac{d^4 k_1}{(2\pi)^4}\frac{d^4 k_2}{(2\pi)^4}
~\frac{1}{2}\tilde{h}(k_2) \left[\mathcal{A}^{(II)}_{\tilde h}\right]\tilde{h}(k_1)\right\}~,
\eea 
where
\bea\label{AhII}
\mathcal{A}^{(II)}_{\tilde h} &=& \frac{1}{9} \left\{M_P^2\left[ \mathcal{P}^{-1}_h(k_1)\right]-(\vec{p}^2\phi_0^2)
\left(1+4\left(\frac{\mathcal{D}_1(\vec{k}_1^2)}{\mathcal{D}_2(\vec{k}_1^2)}\right)+2\left(\frac{\mathcal{D}_1(\vec{k}_1^2)}{\mathcal{D}_2(\vec{k}_1^2)}\right)^2
-\frac{X^2}{6}\frac{\omega_1^2}{\vec{k}_1^2}\right)\right.\nonumber\\
&&-\left.[\bar{\psi}_0(\vec{\gamma}\cdot\vec q)\psi_0]\left(\frac{3}{4}+2\left(\frac{\mathcal{D}_1(\vec{k}_1)}{\mathcal{D}_2(\vec{k}_1)}\right)
-\frac{X^2}{8}\frac{\omega^2_1}{\vec{k}_1^2}\right)\right\}\delta(k_1+k_2)~,
\eea 
and
\bea
\mathcal{P}^{-1}_h(k_1) &=& X \omega_1^2 +\frac{ \vec{k}_1^2\left[1+(3F_1+8F_2)\vec{k}_1^2+(3S_1+8S_2)(\vec{k}_1^2)^2\right]\mathcal{D}_2(\vec{k}_1)
+2\left[\mathcal{D}_1(\vec{k}_1)\right]^2}{-\mathcal{D}_2(\vec{k}_1)}\nn
\mathcal{D}_1(\vec{k}_1) &=& -[\vec{k}_1^2- F_3 (\vec{k}_1^2)^2+S_3 (\vec{k}_1^2)^3]\nn
\mathcal{D}_2(\vec{k}_1) &=& -[\alpha\vec{k}_1^2- F_4 (\vec{k}_1^2)^2+S_4 (\vec{k}_1^2)^3]~.
\eea 
As expected, the IR limit ({\it i.e.} neglecting terms of the form $(\vec{k}^2)^n$ for $n>1$) of the expression (\ref{AhII}) 
leads to the expression (\ref{AhI}), and the dispersion relation for the scalar graviton is given by the identity (\ref{hdrIR}).
The functional integration over $h$ gives
\bea 
&&\exp\left\{\frac{1}{2M_P^2}\left[(\vec p^2 \phi_0^2)+\frac{3}{4}[\bar{\psi}_0(\vec{\gamma}\cdot\vec q)\psi_0]\right]
\mbox{Tr}\left(\frac{\delta(k_1+k_2)}{\mathcal{P}^{-1}_h(k_1)}\right)\right.\\
&&\left.+\frac{2}{M_P^2}\left[(\vec p^2 \phi_0^2)+\frac{1}{2}[\bar{\psi}_0(\vec{\gamma}\cdot\vec q)\psi_0]\right]
\mbox{Tr}\left(\frac{\delta(k_1+k_2)\mathcal{D}_1(\vec{k}_1)}{\mathcal{P}^{-1}_h(k_1) \mathcal{D}_2(\vec{k}_1)}\right)+\frac{(\vec p^2 \phi_0^2)}{M_P^2}
\mbox{Tr}\left(\frac{\delta(k_1+k_2)(\mathcal{D}_1(\vec{k}_1))^2}{\mathcal{P}^{-1}_h(k_1) (\mathcal{D}_2(\vec{k}_1))^2} \right)\right.\nonumber\\
&&\left. -\frac{X^2}{12 M_P^2}\left[(\vec p^2 \phi_0^2) +\frac{ 3}{4}[\bar{\psi}_0(\vec{\gamma}\cdot\vec q)\psi_0]\right]
\mbox{Tr}\left(\frac{\delta(k_1+k_2)\omega_1^2}{\mathcal{P}^{-1}_h(k_1) \vec{k_1}^2}\right)+\cdots\right\}~.\nonumber
\eea
The first three traces above can be written in the following generic form
\be\label{logsc}
\mbox{Tr}\left[\frac{\delta(k_1+k_2)}{\mathcal{P}^{-1}_h(k_1)}\left(\frac{\mathcal{D}_1(\vec{k}_1)}{ \mathcal{D}_2(\vec{k}_1)}\right)^n\right],~~~\mbox{for}~n=0,1,2~,
\ee  
which, after integrating over the frequencies leads to
\bea\label{trn}
&&\mbox{Tr}\left[\frac{\delta(k_1+k_2)}{\mathcal{P}^{-1}_h(k_1)}\left(\frac{\mathcal{D}_1(\vec{k}_1)}{ \mathcal{D}_2(\vec{k}_1)}\right)^n\right]
=\frac{\delta(0)}{2\sqrt{X}}\int \frac{d^3 k}{(2\pi)^3} \left[\mathcal{G}(\vec{k})\left(\frac{D_2(\vec{k})}{D_1(\vec{k})}\right)^{2n}\right]^{-\frac{1}{2}}~,
\eea 
with
\be \label{G}
\mathcal{G}(\vec{k})= (\mathcal{P}_h^{-1} -X \omega^2)=\frac{ \vec{k}^2\left[1+(3F_1+8F_2)\vec{k}^2+(3S_1+8S_2)(\vec{k}^2)^2\right]\mathcal{D}_2(\vec{k})
+2\left[\mathcal{D}_1(\vec{k})\right]^2}{-\mathcal{D}_2(\vec{k})}~.
\ee 
Expanding the terms inside the square brackets in~(\ref{trn}), keeping only terms of at least quartic order in the momentum, as we did in for the tensor field, 
the right-hand side of eq.(\ref{trn}) reduces to
\be\label{3dint} 
\frac{\delta(0)}{2\sqrt{XC_6^{(n)}}}~ \mathcal{I}\left(\frac{C_4^{(n)}}{C^{(n)}_6}\right),
\ee 
with the integral $\mathcal{I}$ given by eq.(\ref{intd}) and, for $n=0,1,2$,
\bea
C_4^{(n)}&=&(c_4^{(n)}/M_{HL}^{2})\\ 
C_6^{(n)}&=&(c_6^{(n)}/M_{HL}^{4})\nn 
c_4^{(0)} &=& -(2 f_1+8f_2)-  \frac{2s_3( f_4 s_3-2 f_3 s_4)}{s_4^2}\nn
c_4^{(1)} &=& 2f_4+\frac{2s_4(2s_1+8s_2)(f_4s_3-f_3 s_4)-s_3s_4^2(2f_1+8f_2)}{s_3^3} \nn
c_4^{(2)} &=& \frac{s_4^2\left[-s_3 s_4^2(2f_1+8f_2) + 4s_4(f_4s_3 -f_3 s_4)(2s_1+8s_2)+2s_3^2(3f_4s_3-2f_3s_4)\right]}{s_3^5}\nn
c_6^{(0)}&=& -(2 s_1+ 8s_2)-\frac{2 s_3^2}{s_4} \nn
c_6^{(1)} &=& -2s_4 -\frac{ (2 s_1+8s_2) s_4^2}{s_3^2}\nn
c_6^{(2)} &=& -2\frac{s_4^3}{s_3^2}-\frac{s_4^4(2s_1+8s_2)}{s_3^4}~.\nonumber 
\eea
Solving this integral with dimensional regularisation using the result~(\ref{Isol}), we obtain
\be\label{convint} 
\mbox{Tr}\left[\frac{\delta(k_1+k_2)}{\mathcal{P}^{-1}_h(k_1)}\left(\frac{\mathcal{D}_1(\vec{k}_1)}{ \mathcal{D}_2(\vec{k}_1)}\right)^n\right] 
= \frac{\delta(0)}{(2\pi)^2\sqrt{X C_6^{(n)}}}\frac{\mu^\epsilon}{\epsilon}+\cdots,
\ee 
where dots represent finite terms.
Finally we need to calculate the following trace
\bea \label{trwk}
\mbox{Tr}\left(\frac{\delta(k_1+k_2)\omega_1^2}{\mathcal{P}^{-1}_h(k_1) \vec{k_1}^2}\right)&=&\delta(0)\int \frac{d^4k}{(2\pi)^4}\frac{\omega^2}{P_h^{-1}(k)\vec{k}^2}\\
&=&\frac{\delta(0)}{ X}\left\{\int \frac{d^4k}{(2\pi)^4}\frac{1}{\vec{k}^2}-\int \frac{d^4k}{(2\pi)^4}\frac{\mathcal{G}(\vec{k})/X}{[\omega^2+\mathcal{G}(\vec{k})/X]\vec{k}^2}\right\}~,\nonumber
\eea 
with $\mathcal{G}$ defined in~(\ref{G}).
The first integral in the last line of~(\ref{trwk}) vanishes with dimensional regularisation. In addition, for the second integral, we integrate over the frequencies to get
\be 
\mbox{Tr}\left(\frac{\delta(k_1+k_2)\omega_1^2}{\mathcal{P}^{-1}_h(k_1) \vec{k_1}^2}\right)=\frac{\delta(0)}{2 X^{3/2}}\int \frac{d^3k}{(2\pi)^3}\frac{\sqrt{\mathcal{G}(\vec{k})}}{\vec{k}^2}~.
\ee 
We then expand $\mathcal{G}(\vec{k})$, for which only the dominant term potentially leads to a divergence
\be 
\mbox{Tr}\left(\frac{\delta(k_1+k_2)\omega_1^2}{\mathcal{P}^{-1}_h(k_1) \vec{k_1}^2}\right)=\frac{\delta(0)\sqrt{C_6^{(0)}}}{2 X^{3/2}}\int \frac{d^3k}{(2\pi)^3} |\vec{k}| ~ + \cdots
\ee 
with dots representing finite terms. The integral vanishes with dimensional regularisation, such that the result is finite.

\vspace{1cm}

\nin{\bf Acknowledgements}
The work  of J. L. is supported by the National Council for Scientific and Technological Development (CNPq - Brazil).

\end{document}